\pdfoutput=1
\documentclass[sigconf]{acmart}

\usepackage{microtype}

\usepackage{bm}

\usepackage{amsmath}

\usepackage{amssymb}

\usepackage{bbm}
\usepackage{graphicx}

\usepackage[ruled,noend]{algorithm2e}

\usepackage{color}

\usepackage{multirow}
\usepackage{booktabs}

\usepackage{makecell}
\usepackage{enumitem}

\renewcommand\footnotetextcopyrightpermission[1]{}

\SetKwInOut{Parameter}{Parameters}

\begin{document}

\title{Order-Disorder: Imitation Adversarial Attacks for Black-box Neural Ranking Models}

\newcommand{\aff}[1]{\texorpdfstring{$^{#1}$}{}}
\author{Jiawei Liu\aff{1}, Yangyang Kang\aff{2}, Di Tang\aff{3}, Kaisong Song\aff{2,4}, Changlong Sun\aff{2},\\Xiaofeng Wang\aff{3}, Wei Lu\aff{1}$^*$, Xiaozhong Liu\aff{5}$^*$}

\affiliation{%
  $^1$ Wuhan University, China ({\{laujames2017,weilu\}@whu.edu.cn})\\
  $^2$ Alibaba Group, China ({\{yangyang.kangyy,kaisong.sks\}@alibaba-inc.com}, changlong.scl@taobao.com)\\
  $^3$ Indiana University Bloomington, USA (tangd@iu.edu, xw7@indiana.edu)\\
  $^4$ Northeastern University, China\\
  $^5$ Worcester Polytechnic Institute, USA (xliu14@wpi.edu)
    \country{}
}

\renewcommand{\shortauthors}{Jiawei Liu et al.}

\begin{abstract}
Neural text ranking models have witnessed significant advancement and are increasingly being deployed in practice. Unfortunately, they also inherit adversarial vulnerabilities of general neural models, which have been detected but remain underexplored by prior studies. Moreover, the inherit adversarial vulnerabilities might be leveraged by blackhat SEO to defeat better-protected search engines. In this study, we propose an imitation adversarial attack on black-box neural passage ranking models. We first show that the target passage ranking model can be transparentized and imitated by enumerating critical queries/candidates and then train a ranking imitation model. Leveraging the ranking imitation model, we can elaborately manipulate the ranking results and transfer the manipulation attack to the target ranking model. For this purpose, we propose an innovative gradient-based attack method, empowered by the pairwise objective function, to generate adversarial triggers, which causes premeditated disorderliness with very few tokens. To equip the trigger camouflages, we add the next sentence prediction loss and the language model fluency constraint to the objective function. Experimental results on passage ranking demonstrate the effectiveness of the ranking imitation attack model and adversarial triggers against various SOTA neural ranking models. Furthermore, various mitigation analyses and human evaluation show the effectiveness of camouflages when facing potential mitigation approaches. To motivate other scholars to further investigate this novel and important problem, we make the experiment data and code publicly available.
\end{abstract}



\maketitle


\section{Introduction}

The two-stage retrieve-then-rerank architecture has proven to be an effective and widely adopted strategy for text ranking tasks \citep{asadi2013effectiveness}. As for the reranker, the trendsetting approach is empowered by the neural model. With the recent accomplishment of pre-trained model, BERT \citep{devlin2019bert} for instance, pre-trained then fine-tuned transformers achieved state-of-the-art performance on numerous text ranking tasks \citep{nogueira2019passage,gao2021coil,pradeep2021expando}. However, neural ranking models may inherit the adversarial vulnerabilities of neural networks \citep{szegedy2013intriguing}, i.e, a small deliberate perturbation (e.g., some pixel variations on an image) could trigger dramatic change in the learning result \citep{zhou2020adversarial}. Such vulnerability has raised dedicated concerns of the robustness and reliability of text ranking systems integrated with neural networks \citep{song2020adversarial}. A trustworthy ranking system should be well aware of the malicious attack, in which deliberate but imperceptible content variations may cause the catastrophic ranking disorder. 


\begin{figure*}[!t]
\centering
  \includegraphics[width=15.5cm]{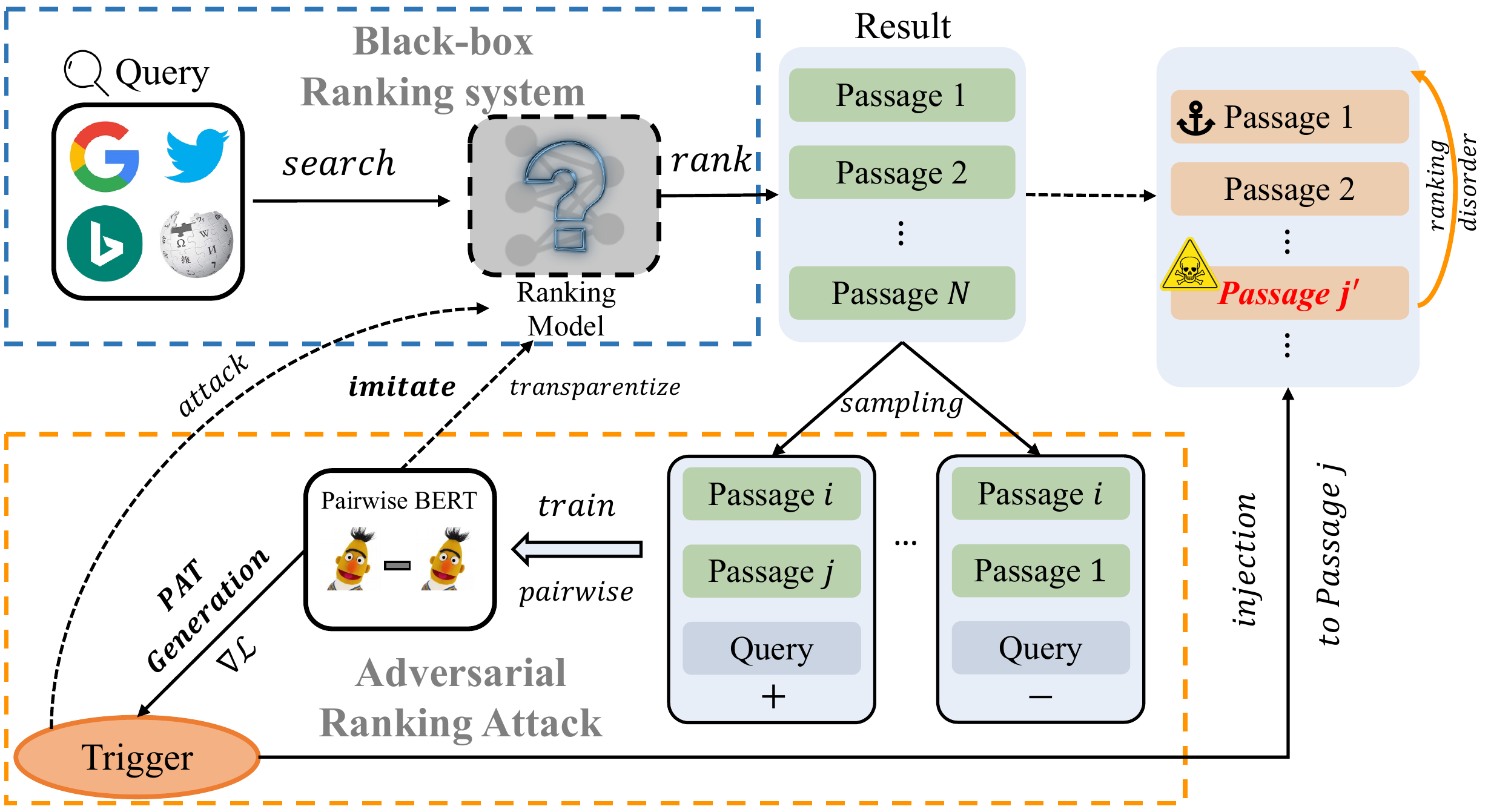}
  \caption{An overview of our imitation adversarial attacks for a black-box neural passage ranking system.}
  \label{overview} 
\end{figure*}

Previous adversarial ranking attacks primarily focus on DNN-based image ranking systems \citep{zhou2020adversarial,zhou2021practical,li2021qair}, whereas the vulnerability of deep neural text ranking remains underexplored. Taking the passage search as an example, a fair ranking system should rank the passage collections according to their semantic relevance to the query. Nevertheless, a malicious content producer may attempt to raise the rank of his/her own passage, e.g., opinion or advertisement, by adding small perturbations to the passage.
Recently, scholars investigated potential adversarial attacks on text classification \citep{ebrahimi2018hotflip,wallace2019universal,song2021universal}, machine translation \citep{belinkov2018synthetic,wallace2020imitation}, and question answering \citep{gan2019improving,tan2020s}, which inspire studies on neural text ranking trustworthiness. For instance, \citet{song2020adversarial} proved that an irrelevant document can be prioritized by inserting a collision text generated based on the target white-box pointwise relevance scoring model. The target model is a BERT classifier fine-tuned with the relevance label of a document to a query. In general, document ranking examines relations among candidate documents in a query context \citep{burges2010ranknet}, and the effectiveness of existing text classification-based adversarial attacks can be compromised. Moreover, when the attack process is coached by gradient, model transparency, e.g., the architecture, hyperparameters, and training data, becomes inevitable \citep{wang2019towards}. Real-world ranking systems, unfortunately, prohibit algorithmic white-box access, which hinders the application of existing attack methods. 

In this study, we propose a novel black-box attack method against neural text ranking systems in terms of adversarial trigger transferability among different neural networks. 
Specifically, we train a ranking imitation model to demystify the target ranking model for knowledge distillation.
The proposed ranking imitation model is a Pairwise BERT ranker, which is trained on triplets \textit{(query, relative positive candidate, relative negative candidate)} sampled from ranking lists of the target model. It does not require the relevance label or score of a document to a query, which is vital to train a pointwise ranker. We prove that the ranking imitation model-inferred candidate relatedness can unveil black-box ranking model vulnerabilities for potential attacks.
Experiments show that the ranking imitation model achieves comparable performance on ranking task with the target text ranking system, i.e., 89.9 vs. 90.1 on TREC DL 2019 \footnote{https://microsoft.github.io/msmarco/TREC-Deep- Learning-2019} MRR@10 and 77.2\% overlap among top-10 candidates. This similarity essentially increases the transferability of our adversarial attack triggers.

In terms of adversarial attack trigger generation, we propose a novel Pairwise Anchor-based Trigger (PAT) generation model by leveraging the pairwise structural information from imitation ranking outcome. The generation process, navigated by the ranking imitation model formulated gradient, can generate the tailored trigger for each candidate passage that carries camouflages to mislead the original ranker. Moreover, to avoid generating nonsensical triggers with high perplexity, which can be easily filtered, we apply the fluency constraint with a language model to the objective. Since the semantic of the trigger that differs significantly from the passage is highly susceptible to be detected, we add the next sentence prediction (NSP) loss between the trigger and the target candidate to equip the trigger camouflages. As a prominent result, our trigger can successfully boost the ranks of nearly 98\% marginal passages (ranked 995-1000) and essentially boost nearly 40\% of them to rank in top-100.
Extensive experiments on passage ranking demonstrate the effectiveness of our adversarial attack trigger generation method and transferability of adversarial triggers.

An example scenario of our attack is to elevate blackhat SEO to defeat better-protected search engines. Our research shows that SOTA embedding-based ranking models, as those used by search engines (Google, Amazon, etc), are vulnerable to blackhat SEO, a well-known cybercrime (e.g., advertising drugs, gambling, and porn \citep{liao2016seeking,he2020creating}). Traditional SEO tricks like keyword (or similar word) stuffing\footnote{https://seranking.com/blog/keyword-stuffing} become less effective with the current trend of moving towards BERT \citep{wu2022prada} + anti-SEO solutions \citep{zhou2009osd}, due to their lack of stealthiness (Section 6.3). Our research, however, shows that ranking manipulation against embedding models can be achieved under the consistency and fluency constraints, even without knowledge of the target, opening the door to stealthy SEO attacks.

The contributions of this paper are threefold:
\begin{enumerate}
    \item[(1)] This is a pioneer transfer-based attack method investigation against black-box neural text ranking systems. By querying the victim ranking system, one can transparentize black-box model via ranking imitation, and generate adversarial attack triggers to disorder the victim ranking system. 
    \item[(2)] We propose an innovative Pairwise Anchor-based Trigger (PAT) generation model with pairwise loss on anchor candidates for camouflaged text ranking attack and manipulation. 
    \item[(3)] We employ three large datasets alone with a variety of ranking models to validate the effectiveness of the proposed model. We make all the experiment data and code publicly available to motivate other scholars to further investigate this novel but important problem\footnote{https://github.com/LauJames/PAT}.
\end{enumerate}

\section{Background and Related Works}
In this section, we briefly introduce the existing works of deep text ranking, adversarial ranking attacks, and model imitation. Then we state the threat model and objectives.

\subsection{Text Ranking}
Classical Text Ranking models mainly rely on exact term matching between query and document text using the bag-of-words framework, such as Boolean Retrieval and BM25 \citep{robertson1994some}. However, this type of method has limited capability of modeling human languages.
Nowadays, due to the efficiency of processing queries and documents, they are still widely used in production systems for first-stage retrieval.
To deal with vocabulary mismatch, continuous vector space representation such word2vec \citep{mikolov2013distributed} and Glove \citep{pennington2014glove} coupled with neural networks to produce soft matching score. These neural ranking models can be generally classified into three classes: representation-based models \citep{huang2013learning,shen2014latent}, interaction-based models \citep{guo2016deep,hofstatter2020interpretable}, and hybrid models \citep{mitra2017learning}.

Deep transformer models pre-trained with language model objectives, represented by BERT \citep{devlin2019bert}, have made a huge impact on neural text ranking. \citet{nogueira2019passage} are the first to demonstrate the effectiveness of BERT in text ranking task. After that, more and more pre-trained transformer LM models fine-tuned on the specific corpus achieve state-of-the-art in text ranking with significant performance improvement \citep{dai2019deeper,gao2021coil,pradeep2021expando}. In the Deep Learning Track at TREC 2019 \citep{craswell2020overview}, analysis of the results showed that, BERT-based models achieved substantially higher effectiveness than ``pre-BERT'' models, across implementations by different teams \citep{yates2021pretrained}. To lower the query latency and make the ranking model feasible for production deployment, \citet{hofstatter2020improving} adapt knowledge distillation to neural ranking models.
However, the performance improvement of BERT-based text ranking models also inherited the vulnerabilities of neural networks, which have been detected \citep{song2020adversarial} but still remain under-explored by prior studies.

\subsection{Adversarial Ranking Attack}
\citet{szegedy2013intriguing} find that DNN is susceptible to small adversarial perturbation added to inputs, which leads to misbehavior. Following this finding, a series of researches and applications about adversarial attacks have emerged in text classification \citep{ebrahimi2018hotflip,wallace2019universal,song2021universal}, machine translation \citep{belinkov2018synthetic,wallace2020imitation}, and question answering \citep{gan2019improving,tan2020s}. Different from classification tasks where texts are predicted independently, the rank of one candidate is usually related to the query as well as other candidates for ranking tasks. Since the ranking result is determined by the relative relations among candidates and queries, it is essential to take the relations into consideration for a qualified ranking model. As a result, for text ranking scenarios, the existing adversarial text classification attack methods are incompatible.

For ranking systems, the risk of malicious user manipulating the target ranking always exists \citep{goren2018ranking}.
Previous adversarial ranking attacks primarily focus on DNN-based image ranking systems \citep{zhou2020adversarial,li2021qair,wang2020transferable}. Likewise, the performance improvement of BERT-based document ranking models \citep{yang2019simple} also inherit the vulnerabilities of neural networks \citep{song2020adversarial}. The existence of aforementioned works in NLP and image ranking inspired the attack against deep text ranking models, it is still insufficiently explored. Although there are several research works explore the attack against neural text ranking models, they focus on document ranking \citep{song2020adversarial} and are under white-box setting \citep{raval2020one,goren2020ranking,song2022trattack,wu2021neural}.


Considering a majority of real-world text ranking systems do not allow white-box access, recently, \citet{wu2022prada} propose a pseudo relevance-based adversarial ranking attack method, which substitutes words to promote the target document in rankings. Differently, we introduce the transfer-based black-box attack and train a ranking imitation model on triplets sampled from ranking lists of the target ranking model. Leveraging the similarity of the ranking imitation model, we can elaborately manipulate the ranking results and transfer the manipulation attack to the target ranking model. Then we propose our trigger generation method, empowered by the pairwise objective function, to generate adversarial triggers, which causes premeditated disorderliness with very few tokens.

\subsection{Model Imitation}
Model imitation is closely related to model distillation \citep{hinton2015distilling} and extraction (or stealing) \citep{krishna2019thieves}. Model distillation aims to train a student model to imitate the predictions of a teacher. \citet{hofstatter2020improving} investigate the Margin-MSE knowledge distillation across different architectures to improve reranking effectiveness. Recent works have shown that backdoors can persist in the student model if the teacher model is infected with a Trojan \citep{wang2020backdoor,yao2019latent,wallace2020imitation}. However, most of the model distillation methods utilize the logits or scores of the teacher model, which are unavailable and prohibited in black-box setting.  Thus, ranking imitation models cannot be optimized by distribution matching losses commonly used in knowledge distillation, such as Margin-MSE and KL divergence. In terms of model extraction or stealing, it differs from distillation because the training data of the victim (teacher) is unknown, which causes queries for the victim to be out-of-domain.

Despite the above-mentioned challenges, previous works show that model extraction or stealing is possible for text classification \citep{pal2019framework}, reading comprehension, natural language inference \citep{krishna2019thieves}, and machine translation \citep{wallace2020imitation}. In this paper, we extend these results to text ranking. Generally speaking, absolute positive labels or scores are required to train a pointwise ranker. However, this information is inaccessible for real-world ranking systems. We introduce the Pairwise BERT ranker to imitate a black-box ranking model. The Pairwise BERT ranker is trained on triplets \textit{(query, relative positive candidate, relative negative candidate)} sampled from ranking lists of the target model.

\section{Threat Model}
\label{threat}

Given a textual query $\boldsymbol{q}$ and candidate passages set $\boldsymbol{P}=\{\boldsymbol{p}_1,\boldsymbol{p}_2,...,\boldsymbol{p}_{k}\}$, the ranking model calculates a score $s(\boldsymbol{q}, \boldsymbol{p}_i)$ with respect to a candidate passage $\boldsymbol{p}_i \in \boldsymbol{P}$ and query $\boldsymbol{q}$, and generates ranking list, i.e., $\boldsymbol{p}_1 \succ \boldsymbol{p}_2 \succ \cdots \succ \boldsymbol{p}_k$ if $s(\boldsymbol{q}, \boldsymbol{p}_1) > s(\boldsymbol{q}, \boldsymbol{p}_2) > \cdots > s(\boldsymbol{q}, \boldsymbol{p}_k)$.

\textbf{Objective of the Adversary.}
The adversarial text ranking attack aims to find an optimized adversarial trigger (text snippet) which leads to the deliberate ranking disorder. For example, given $s(\boldsymbol{q}, \boldsymbol{p}_i) > s(\boldsymbol{q}, \boldsymbol{p}_j)$, trigger $\boldsymbol{t}$ injected into $\boldsymbol{p}_j$ (i.e., ${\boldsymbol{p}'}_j=[\boldsymbol{t}; \boldsymbol{p}_j]$, where $;$ denotes concatenation) can flip ranking result, $s(\boldsymbol{q}, {\boldsymbol{p}}_i) < s(\boldsymbol{q}, \boldsymbol{p}'_j)$. To make the attack as stealthy as possible, the trigger should be avoided to be nonsensical with high perplexity. Meanwhile, the passage injected with the trigger should be semantically consistent with the original passage.

\textbf{Capabilities of the Adversary.} 
We focus on the black-box attack setting where the adversary has no knowledge of the target model architecture, training data, and the score function. Accessing ranking result list $\boldsymbol{P}$ produced by the target victim model with the query $\boldsymbol{q}$ is one of the capabilities assumed in our threat model. Moreover, we also assume that there is possibility to get a small part of the original training data \citep{carlini2021extracting} of the ranking model. The adversary could train a ranking imitation model to transparentize the target ranking model with a series of queries $\boldsymbol{q}_M$, ranking results $\boldsymbol{P}_M$, and possible original training data, where $M$ is number of query. Based on the findings of Pseudo Relevance Feedback in information retrieval \citep{dehghani2017neural}, the ranking imitation model is trained with weak supervision manner to imitate the target ranking model. Also, the ranking imitation model generated adversarial trigger can be transferred to the victim model via gradient-based search \citep{wallace2020imitation}.

\textbf{Target Attack Model.} We select BERT-base \citep{devlin2019bert,nogueira2019passage} and MiniLM-L-12 \citep{wang2020minilm} fine-tuned on passage ranking dataset as representative passage ranking models to study for our attack. Fine-tuning pre-trained BERT that adopts query and passage concatenation architecture is the common way now and achieves previous SOTA in text ranking with huge performance leap \citep{nogueira2019passage}. As for the fine-tuned MiniLM, it is the cross-encoder architecture \citep{luan2020sparse}, which is another widely adopted architecture, and adopts MiniLM as encoder. It achieves highly ranked performance and also is previous SOTA.

\section{Methodology}
\label{method}
In this section, according to the formulated black-box ranking attack problem in section \ref{threat}, we elaborate the proposed attack method.

\subsection{Overview}
The whole attack pipeline is depicted in Figure \ref{overview}, which carries two key phases: black-box ranking model imitation phase and Pairwise Anchor-based Trigger (PAT) generation phase.

\begin{algorithm}[!t] 
\caption{Imitation adversarial attacks for black-box text ranking models}  
\label{alg:overview}  
\DontPrintSemicolon
\LinesNumbered
\KwIn{target black-box ranking model $\boldsymbol{O}$, \textit{Pairwise BERT} $\boldsymbol{R}$, query collection $\boldsymbol{Q}$,  target query $\boldsymbol{q}$, target passage $\boldsymbol{p}$, anchor passage $\mathcal{A}$, language model $g$, vocabulary $\mathcal{V}$}
\Parameter{sample top-n size $N$, trigger length $L$, epochs $E$, coefficient $\epsilon$, temperature $\tau$, step size $\alpha$, beam size $B$, beam top-n size $U$}
\KwOut{an adversarial trigger $\boldsymbol{t}$}
\SetKwProg{Proc}{Phase}{}{}
\Proc{1. Black-box Ranking Model Imitation}{
    INIT: Dataset $\mathcal{D} \gets \{\}$ \\
    \For{$\boldsymbol{q}_m \in \boldsymbol{Q}$}{
         rank list $\boldsymbol{P}_m \gets$ query $\boldsymbol{O}$ with $\boldsymbol{q}_m$ \\
         \For{$\boldsymbol{p}_i,\boldsymbol{p}_j \in \text{top-N}(\boldsymbol{P}_m) \& Rank(\boldsymbol{p}_i)>Rank(\boldsymbol{p}_j)$}{
            $\mathcal{D} \gets positive, [\boldsymbol{q}_m; \boldsymbol{p}_i; \boldsymbol{p}_j]$ \\
            $\mathcal{D} \gets negative, [\boldsymbol{q}_m; \boldsymbol{p}_j; \boldsymbol{p}_i]$ \\
            // Reverse $\boldsymbol{p}_i$ and $\boldsymbol{p}_j$ to get the negative triple.\\
         }
    }
    Train the ranking imitation model $\boldsymbol{R}$ on $\mathcal{D}$ using Eq \ref{eq:loss_imitate}.\\
    \Return{$\boldsymbol{R}$}\
}
\SetKwFunction{FMain}{LogitsPerturbation}
\SetKwProg{Fn}{Function}{:}{}
  
\Proc{2. Pairwise Anchor-based Trigger Generation}{
    INIT: $\boldsymbol{\Theta} = [\boldsymbol{\theta}_1,\dots,\boldsymbol{\theta}_L]$, where $\boldsymbol{\theta}_l\in\mathbb{R}^{|\mathcal{V}|} $ \& $\sim \mathcal{N}(0,1)$ \\
    Top-n similar words and anchor passage words $\mathcal{V}_{sim}$\\
    \Fn{\FMain{$\ell$, $\boldsymbol{t}_{1:l-1}$}}{
        $\boldsymbol{\rho} \gets \boldsymbol{0} \in\mathbb{R}^{|\mathcal{V}|}\&$ \\
        $w + \epsilon$ for $w$ in $\boldsymbol{\rho}$ if $ w\in \mathcal{V}_{sim}$ \\
        \For{epoch$\gets 1$ to $E$}{
            $\boldsymbol{\tilde{t}}_{l}\gets\text{softmax}((\boldsymbol{\ell}+ \boldsymbol{\rho}_l)/\tau)$\\
            $\boldsymbol{\rho}\gets\boldsymbol{\rho}$ - $\alpha \cdot {\nabla_{\boldsymbol{\rho}}}\mathcal{L}_{\boldsymbol{R}}(\boldsymbol{q},\boldsymbol{t}_{1:l-1};\tilde{\boldsymbol{t}}_l,\mathcal{A})$\\
        }
        \Return{$\boldsymbol{\theta}=\boldsymbol{\ell} + \boldsymbol{\rho}$}
    }
    $\mathcal{B} \gets$ $B$ replicates of <BOS> token \\
    \For{$l\gets$ 1 to $L$}{
        Beam score matrix $\boldsymbol{S}_l \gets \boldsymbol{0}\in\mathbb{R}^{B\times{U}}$ \\
        \For{ each beam $\boldsymbol{t}_{1:l-1} \in \mathcal{B}$}{
            $\boldsymbol{\ell}_l \gets \text{next token logits from LM } g(\boldsymbol{t}_{1:l-1})$\\
            $\boldsymbol{\theta}_l \gets$ \FMain{$\boldsymbol{\ell}_l,\boldsymbol{t}_{1:l-1}$}\\
            \For{$w \in$ top-U of $\boldsymbol{\theta}_l$}{
                $\boldsymbol{S}_l[\boldsymbol{t}_{1:l-1}, w] \gets$ objective score from Eq \ref{objective}.
            }
            $\mathcal{B} \gets \{\boldsymbol{t}_{1:l-1};w|(\boldsymbol{t}_{1:l-1}, w)\in \text{top-}B\text{ of } \boldsymbol{S}_l\}$
        }
    }
    \Return{$\boldsymbol{t}=\text{argmax }\mathcal{B}$}\
}

\end{algorithm}

In \textit{Phase 1}, to enable the adversary to train a ranking imitation model, which can mimic the ranking list of the black-box victim ranking model without a real labeled dataset, we utilize the relative relevance information among the ranking result list of the victim ranking model to construct a synthetic dataset. Specifically, we first sample triples from the ranking result list by querying the black-box victim ranking model. The triple is composed by one query, one relative positive candidate, and one relative negative candidate, where the relative positive candidate is ranked ahead of relative negative candidate. Then we can train an ranking imitation model, which adopts pairwise encoder architecture, i.e., pairwise BERT, with those sampled triples.

In \textit{Phase 2}, the adversary utilizes the ranking imitation model to craft adversarial triggers. By inserting the trigger into one irrelevant passage, the rank of the passage, produced by the target victim ranking model, should be boosted due to the transferability of the ranking imitation model. Moreover, we propose a Pairwise Anchor Trigger generation method to make full use of the pairwise structural information of the ranking imitation model.


Our proposed imitation adversarial attack for black-box neural text ranking attack is outlined in pseudo Algorithm \ref{alg:overview}.


\subsection{Black-box Ranking Model Imitation}
\label{method_imitate} 
Practically, the adversary has no knowledge of the target model architecture, score function, and ground truth label information. Nevertheless, the orientation of transparentization is to train a ranking imitation model that substitutes and achieves comparable performance to the victim model. Since we have query access to the ranking result list, which consists of ranks of candidates, based on the findings of Pseudo Relevance Feedback in information retrieval, we can train a ranking imitation model by sampling triples from the ranking result list to imitate the target ranking model. To enhance the transferability of adversarial triggers, the ranking imitation model should be functionally similar to the victim model \citep{wallace2020imitation}. In this work, we measure the model similarity by the similarity of ranked lists.

Suppose we get the rank list of $K$ passages $\boldsymbol{P}_m=[\boldsymbol{p}_1,\dots,\boldsymbol{p}_K]$ with respect to the query $\boldsymbol{q}_m \in \boldsymbol{Q}=\{\boldsymbol{q}_1,\dots,\boldsymbol{q}_{M} \}$ from the victim model, we sample triples $[\boldsymbol{q}_m; \boldsymbol{p}_i; \boldsymbol{p}_j]$ from the top $N$ candidate passages with label $\boldsymbol{1}$,
where $\boldsymbol{p}_i$ is ranked ahead of $\boldsymbol{p}_j$ ($i\ne j$) and $1\leq i,j \leq N \leq K$. 
The sampled triples \textit{(query, relative positive candidate, relative negative candidate)} carry \textit{pseudo hard examples} and the relative relevance information among the top-$N$ results of the victim model. Thus, by applying our sampling strategy, we can get the ranking model imitation dataset $\mathcal{D}$ and learn an imitation ranking model efficiently. Moreover, it does not require the specific relevance label or score of the document to the query, which is necessary to train a pointwise ranker. The ranking imitation model-inferred candidate relatedness could unveil black-box ranking model vulnerabilities.

We then train a ranking imitation model \textit{Pairwise BERT} based on sampled triples to substitute the victim. 
In the Pairwise BERT ranking model, the query and a pair of passages are concatenated with [SEP] and [CLS] tokens. We utilize the concatenation paradigm as our base architecture, as it represents the current state-of-the-art in terms of rerank effectiveness. The scores are computed by a single linear layer $\boldsymbol{W} \in \mathbb{R}^{768 \times 2}$. The whole process can be formulated as below:
\begin{align}
    \boldsymbol{s}_{i} = \text{BERT}([\text{CLS};\boldsymbol{q}_m;\text{SEP};\boldsymbol{p}_i;\text{SEP}])*\boldsymbol{W}\\
    \boldsymbol{s}_{j} = \text{BERT}([\text{CLS};\boldsymbol{q}_m;\text{SEP};\boldsymbol{p}_j;\text{SEP}])*\boldsymbol{W}
\end{align}
where $\boldsymbol{s}_{i}$ and $\boldsymbol{s}_{j} \in \mathbb{R}^2$ represent the positive scorer (a.k.a., $\boldsymbol{s}_{pos}$) and negative scorer, respectively.

The Pairwise BERT actually train one BERT model applied on two concatenated sequences. The loss is computed as follow:
\begin{align}
    \mathcal{L}_{R}(\boldsymbol{q}_m,\boldsymbol{p}_i,\boldsymbol{p}_j) = -{\boldsymbol{y}}_{m,i,j} \log(\text{softmax}(\boldsymbol{s}_{i} - \boldsymbol{s}_{j}))
    \label{eq:loss_imitate}
\end{align}
where $\boldsymbol{y}_{m,i,j} \in \mathbb{R}^2$ denotes the one-hot label of $[\boldsymbol{q}_m; \boldsymbol{p}_i; \boldsymbol{p}_j]$. To balance the label distribution, we get the negative label of the triples by exchanging the positive passage and the negative passage. We use backpropagation to compute the parameter gradients with Adam \citep{kingma2014adam} optimizer.
\textbf{During inference, we get the last dimension of $\boldsymbol{s}_{i}$ as the score of the passage with respect to the query.}

\label{trigger}
\begin{figure}[!t]
\centering
  \includegraphics[width=7.5cm]{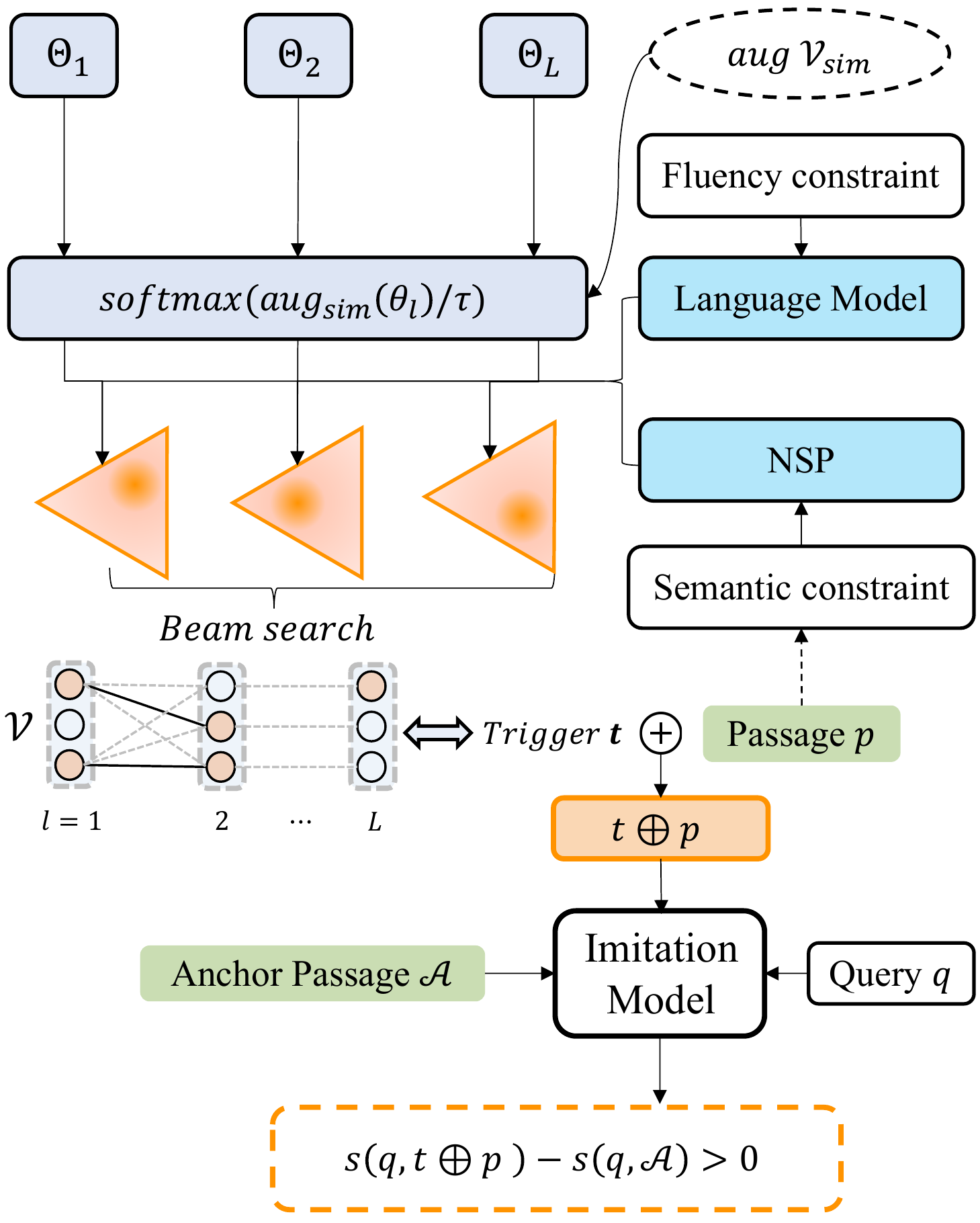}
  \caption{Overview of Pairwise Anchor-based Trigger (PAT) generation.}
  \label{pat} 
\end{figure}

\subsection{Pairwise Anchor-based Trigger (PAT) Generation}
\label{subsection:pat}

With \textit{Pairwise BERT} ranking imitation model, we can create adversarial triggers for black-box ranking models by leveraging the transferability of adversarial examples \citep{papernot2017practical}. To make full use of the pairwise structural information, we propose a Pairwise Anchor-based Trigger (PAT) generation method (Figure \ref{pat}). Given a target query sentence $\boldsymbol{q}$, a target candidate passage $\boldsymbol{p}$, and the top candidate passage $\mathcal{A}$ (anchor), we aim to generate a trigger $\boldsymbol{t}$ for the ranking imitation model with its pairwise loss function $\mathcal{L}_{R}$.
The anchor is used to guide the trigger generation, i.e.,:
\begin{align}
\label{loss_main_attack}
    \operatorname{min} \mathcal{L}_{R}(\boldsymbol{q},\boldsymbol{t} \oplus \boldsymbol{p}, \mathcal{A})
\end{align}
where $\oplus$ denotes injection, e.g., adding the triggers in front of the target passage. We use a gradient-based search to generate a fixed length trigger for the target input.

Inspired by the generation of adversarial semantic collision \citep{song2020adversarial}, we also first utilize gradient optimization with relaxation to find a soft continuous representation of a trigger. Then we combine beam search with score function of ranking imitation model to find the trigger. Label smoothing can be applied at each iteration to optimize the objective.
Formally, we have the vocabulary $\mathcal{V}$ of the model and the length $L$ of the trigger. Each word of trigger at position $l$ is represented as continuous logit vector $\boldsymbol{\theta}_l \in \mathbb{R}^{|\mathcal{V}|}$.
To enhance the attack efficiency, we apply an augmentation function to $\boldsymbol{\theta}_l$. 
Specifically, the logit value corresponding to similar words $\mathcal{V}_{sim}$ will be augmented by adding $\epsilon$, denoted as $\operatorname{aug}_{sim}$.
Note that similar vocabulary consists of tokens from anchor passage and tokens selected by the ranking imitation model with the query and the model vocabulary. Then they are softly selected to input words:
\begin{align}
    \tilde{\boldsymbol{t}}_l = \operatorname{softmax}(\operatorname{aug}_{sim}(\boldsymbol{\theta}_l)/\tau)
\label{relax}
\end{align}
where $\tau$ is a temperature scalar, which controls the sharpness of word selection probability. We can get the probability of each word the vocabulary $\mathcal{V}$ by applying the softmax function on the continuous variable $\boldsymbol{\theta}_l$. 


To avoid generating nonsensical triggers with high perplexity, which can be easily filtered, we add the fluency constraint with a language model (LM) $g$. The semantic of the trigger that differs significantly from the passage is highly susceptible to be detected. To equip the trigger camouflages, we add the next sentence prediction (NSP) model. Specifically, we apply the sequential optimization to the trigger by LM decoding with the joint search on the ranking score $\boldsymbol{s}_{pos}$, LM likehood, and next sentence prediction score $\boldsymbol{f}_{nsp}$. At each time step $l$, we need to maximize:
\begin{equation}
\begin{split}
\max_{w \in \mathcal{V}}   \boldsymbol{s}_{pos}\left(\boldsymbol{q}, \boldsymbol{t}_{1: l-1}; w\right) 
 + \lambda_{1} \cdot \log P_{g}\left(\boldsymbol{t}_{1: l-1};w \right) \\
 + \lambda_{2} \cdot \boldsymbol{f}_{nsp}\left(\boldsymbol{p},\boldsymbol{t}_{1: l-1};w \right)
\end{split}
\label{objective}
\end{equation}
where $\boldsymbol{t}_{1: l-1}$ is the searched beam sequence before $l$. $\boldsymbol{s}_{pos}$ is the positive scorer of the ranking imitation model. $\boldsymbol{f}_{nsp}$ is the probability that $\boldsymbol{t}_{1: l-1};w$ is the continuation of $\boldsymbol{p}$. ${\lambda}_1$ and ${\lambda}_2$ are the hyperparameters that controls the trigger generation constraint strength of the semantic consistency and fluency, respectively.

Inspired by the works of \citet{dathathri2019plug} and \citet{song2020adversarial}, we combine the logits to take ranking similarity and semantic consistency into account. Suppose we get the next-token logits $\boldsymbol{\ell}_l$ generated by LM $g$ at step $l$, then we optimize from the current context to find an update that favors tokens maximizing the ranking score and semantic consistency. Let $\boldsymbol{\theta}_{l}=\boldsymbol{\ell}_l+\boldsymbol{\rho}_l$, where $\boldsymbol{\rho}_l \in \mathbb{R}^{|\mathcal{V}|}$ is a perturbation variable by sampling from $\mathcal{N}(0,1)$. Then we optimize the relaxed pairwise ranking loss  objective $\min_{\boldsymbol{\rho}_{l}}\mathcal{L}_{R}\left(\boldsymbol{q}, \boldsymbol{t}_{1: l-1}; \tilde{\boldsymbol{t}}_{l},\mathcal{A} \right)$for several epoch steps, where $\boldsymbol{\tilde{t}}_l$ is the relaxed soft word from equation \ref{relax}. We minimize the pairwise ranking loss stochastically using Adam \citep{kingma2014adam} optimizer. This operation searches the next-token prediction distribution combining with the perturbed logits to favor words that are likely to boost the rank of a passage.

After perturbation at each time step $l$, we can get the top-n candidate words in $\tilde{t}_l$. With the subset of tokens $w$, we can calculate the ranking score $\boldsymbol{s}_{pos}\left(\boldsymbol{q}, \boldsymbol{t}_{1: l-1}; w\right)$ and the semantic consistency score $\boldsymbol{f}_{nsp}\left(\boldsymbol{p},\boldsymbol{t}_{1: l-1};w \right)$. Given the current beam context, these words $w$ are likely under the LM prediction. Each top-n word in a beam path is assign a score as in equation \ref{objective}. We update the beams with the top-scored words. Since we adopt the hard word at each optimization step, the method can generate a natural-looking ranking attack trigger with semantic consistency.




\section{Imitating Black-box Ranking Models}

In this section, we present the imitation experiment results with multiple datasets. For imitations and attacks, we implement our approach with PyTorch \footnote{https://pytorch.org/} and the HuggingFace Transformers library \citep{wolf2019huggingface}.

\subsection{Datasets}

We mainly adopt three passage ranking datasets to study the performance on different domains:

\textbf{MSMARCO DEV}. The MAchine Reading COmprehension (MSMARCO) dataset \citep{nguyen2016ms} is based on sampled real users' Bing queries. The corpus is initially constructed by retrieving the top-10 passages from the Bing search engine and then annotated. Relevance labels are sparsely-judged and derived from what passages are marked as having the answer to the query. The full training set contains approximately 400M tuples of a query, relevant and non-relevant passages. The development set (MSMARCO DEV) of passage reranking contains 6,980 queries, each paired with the top 1,000 passages retrieved with BM25 from the MSMARCO corpus.

\textbf{TREC DL 2019}. The passage ranking task of TREC Deep Learning Track 2019 (TREC DL 2019) dataset \citep{craswell2020overview} is similar to MSMARCO. But the official evaluation set provides 200 queries, 43 of which are manually and densely judged by NIST assessors with graded relevance labels. For queries with many relevant passages, additional passages were judged as a graded relevance label in \{\textit{Perfectly Relevant (3), Highly Relevant (2), or , Related (1), Irrelevant (0) }\} \citep{craswell2020overview}. Note that, we utilize the officially labeled 43 queries to evaluate the ranking model imitation performance, to keep the same with previous works adopting TREC DL 2019 dataset. As for evaluating the performance of ranking attack, we utilize the random sampled 100 queries and select target passages ranked by the victim model.

\textbf{NQ}. The Natural Question (NQ) dataset is created by \citet{kwiatkowski2019natural} for open-domain QA. The original dataset contains 300K questions collected from Google search logs. We adopt the new version of NQ \citep{karpukhin2020dense} further selected and processed the Wikipedia articles as the collection of passages. This version of NQ dataset contains more than 60K questions, 21M passages, and corresponding negative samples. The NQ dataset is adopted as out-of-domain dataset to firstly train a decent passage ranking model, which is then used to further imitate target victim ranking model.


\textbf{TREC MB 2014}. The TREC Microblog 2014 (TREC MB 2014) dataset \citep{lin2014overview} uses tweets collection to evaluate the model for social media search. It contains 50 queries. Following the experimental procedure of prior works \citep{rao2019multi,lin2014overview}, we evaluate
models in a reranking task, using the top-1000 tweets preprocessed by \citet{rao2019multi}\footnote{https://github.com/jinfengr/neural-tweet-search}.

We sample imitation data from the results on MSMARCO DEV of publicly available models. Models are also tested on TREC MB 2014, which is an \textit{out-of-domain} dataset for these models, to demonstrate the \textit{zero-shot ranking} performance. We also test the ranking performance and functionality similarity (inter agreement) between the victim and ranking imitation models on TREC DL 2019.

Note that the ground truth relevance labels of the open-source dataset are manually assigned, while the training data for our pairwise ranking imitation model were labeled through random sampling from the ranking list of the target model, which does not require white-box access to the target and manual work.

\subsection{Baselines and Target Models}
We compare our ranking imitation models with the following most widely used and state-of-the-art text ranking models. 

1) \textbf{BM25} \citep{robertson1994some}:
It is a classic exact lexical match algorithm with high efficiency that is widely adopted for the first stage retrieval (recall). 
We adopt the tuned parameters and results from Anserini\footnote{https://github.com/castorini/anserini}.

2) \textbf{TK} \citep{hofstatter2020interpretable}: The Transformer-Kernel (TK) model is not based on BERT pre-training, but rather uses shallow Transformers and pre-trained word embeddings. We directly copy the reported numbers from \citet{hofstatter2020improving}.

3) \textbf{BERT-Base} and \textbf{BERT-Large} \citep{nogueira2019passage}: 
It is the common way of utilizing the pre-trained BERT \citep{devlin2019bert} model in reranking scenario that concatenates query and passage input sequences and achieves previous state-of-the-art in text ranking with huge performance leap.
We utilize the fine-tuned publicly available BERT-Base and BERT-Large for passage ranking task from NBoost\footnote{https://github.com/koursaros-ai/nboost}.

4) \textbf{DistilBERT$_{CAT}$} \citep{hofstatter2020improving}: It is a DistilBERT concatenation model trained with Margin-MSE loss that uses vanilla, mono, and concatenated DistilBERT rerankers as teachers and ensembled on MSMARCO passage dataset.
We adopt the fine-tuned model published by \citet{hofstatter2020improving}\footnote{https://huggingface.co/sebastian-hofstaetter/distilber-cat-margin\_mse-T2-msmarco}.

5) \textbf{MiniLM} \citep{wang2020minilm}: This model utilizes the fine-tuned MiniLM as encoder and is the cross-encoder architecture overall \citep{luan2020sparse}. We adopt the MiniLM-L-12 version, which achieves highly ranked performance on MSMARCO and TREC DL 2019\footnote{https://huggingface.co/cross-encoder/ms-marco-MiniLM-L-12-v2}.

6) \textbf{Pairwise BERT (ours)}: Model details have been explained in Section \ref{method}. Practically in this paper, we adopt the BERT-base-uncased model as the encoder and the pairwise pattern to train the ranking imitation model.

\subsection{Research Questions and Experimental Settings} 
In practice, the adversary will not know the architecture or original training dataset of the victim ranking model. In this paper, we study the effect of the following variables on ranking model imitation:

\begin{itemize}[leftmargin=*]
\item \textit{RQ1: How the sampling strategy affects the ranking model imitation performance?} Ranking imitation model training triples $\mathcal{D}$ are constructed by the query, top-$N$ candidates and other candidates pairs randomly selected from the ranking results of MiniLM on MSMARCO DEV, where $N \in\{15,20,25,29\}$. For fair comparison, we restrict the total numbers of sampled triples as the same scale. Specifically, we explore two scales, which are approximately 0.4K (top-15 + other 19 candidates; top-20 + other 10 candidates; top-25 + other 4 candidates; top-29 candidates) and 1K (top-15 + other 59 candidates; top-20 + other 40 candidates; top-25 + other 28 candidates; top-29 + other 20 candidates) sampled pairs per query, respectively. The sampled pairs are mainly from top-K results. However, for each query, top-K may not contain enough hard pairs, e.g., $C^{20}_{2}$ is smaller than 0.4K. To keep the same number of total sampled instances, we also sample some passages not in the top-K results as $\boldsymbol{p}_j$. The relative relevance (in Section \ref{method_imitate}) between $\boldsymbol{p}_i$ and $\boldsymbol{p}_j$ still holds.
\item \textit{RQ2: Can pre-trained Pairwise BERT model improve the performance of ranking model imitation?} Given a specific sampled triples $\mathcal{D}$ for training the ranking imitation model, we use different datasets, which are MSMARCO (In-Domain, ID), NQ (Out-Of-Domain, OOD), and Null dataset (Zero), to train the Pairwise-BERT firstly to get a decent ranker. Specifically, the MSMARCO (ID) used in ranking model imitation task is a small random sampled MSMARCO training dataset (2.56M), while other victim models use triples ranging from 12.8M to 160M. Both of the rankers firstly trained on ID and OOD datasets are tuned with the sampled triples $\mathcal{D}$ to get the final ranking imitation models. Whereas the \textit{Zero Imitate} imitates other victims by straightly training on sampled victims' triples without any pre-training process.
\item \textit{RQ3: How do the differences in architecture, hyperparameters and pre-trained corpus between the ranking imitation model and the victim model affect the imitation performance?} To construct a calibration value of ranking model imitation effectiveness, we firstly train the \textit{Pairwise BERT(ID)} in Table \ref{tab:extract} and use the \textit{all same} Pairwise BERT, i.e., \textit{Imitate ID's Triples}, to imitate \textit{Pairwise BERT(ID)} based on the predicted triples of it. Then we choose \textit{MiniLM} and \textit{BERT-Large} as victim ranking models to study the effect of different and similar architectures on ranking model imitation, respectively. Note that BERT-Large is trained on MSMARCO training dataset and utilizes \textit{similar} but \textit{mismatched} architectures and parameters. Whereas MiniLM utilizes \textit{different} pre-train data, architecture and parameters\footnote{Although it uses transformer layers, the distilled knowledge, learned knowledge, and parameters are different with other BERT-based models. Hence, we perceive it as the victim with different pre-train data, architectures, and parameters.}.
\end{itemize}


Note that we do not adopt other models based on term-frequency or word2vec, because their ranking results are not as effective as the transformer-based contextualized models. They may not be as widely used as the deep transformer-based contextualized models in reranking scenario, facing relatively small threats. Since our method is mainly based on ranking results, it can be easily generalized to other types of ranking models. We will further explore this problem in the future.


\begin{table*}[t]
  \centering
  \renewcommand\tabcolsep{2pt}
  \caption{ Ranking performance (\%) of imitating MiniLM on MSMARCO DEV and TREC DL 2019 datasets using different triple sampling strategies. Note that the performance of MiniLM is irrelevant to the number of sampled pairs per query. We add it to the first row of the table to facilitate a comparison of imitation performance. \textbf{\textit{Bold}} shows the best imitation performance of a model under different sampling strategies. ``-'' means not applicable.}
  \resizebox{1\textwidth}{!}{
    \begin{tabular}{c|c|cc|cccc|cc|cccc}
    \toprule
    \multirow{3}[6]{*}{\textbf{Model}} & \multirow{3}[6]{*}{\textbf{\makecell[c]{Sampling\\Strategy}}} & \multicolumn{6}{c|}{\textbf{$\approx$0.4K sampled pairs per query}} & \multicolumn{6}{c}{\textbf{$\approx$1K sampled pairs per query}} \\
\cmidrule{3-14}          &       & \multicolumn{2}{c|}{\textbf{MSMARCO}} & \multicolumn{4}{c|}{\textbf{TREC DL 2019}} & \multicolumn{2}{c|}{\textbf{MSMARCO}} & \multicolumn{4}{c}{\textbf{TREC DL 2019}} \\
\cmidrule{3-14}          &       & \textbf{MRR@10} & \textbf{NDCG@10} & \textbf{MRR@10} & \textbf{NDCG@10} & \textbf{Inter@10} & \textbf{RBO@1K} & \textbf{MRR@10} & \textbf{NDCG@10} & \textbf{MRR@10} & \textbf{NDCG@10} & \textbf{Inter@10} & \textbf{RBO@1K} \\
    \midrule
    \midrule
    MiniLM (V2) & \textit{Target} & \textit{39.7} & \textit{45.6} & \textit{90.1} & \textit{74.3} & \textit{-} & \textit{-} & \textit{39.7} & \textit{45.6} & \textit{90.1} & \textit{74.3} & \textit{-} & \textit{-} \\
    \midrule
    \multirow{4}[2]{*}{Zero Imitate V2} & Top15+others & 37.1  & 43.3  & 85.3  & 67.8  & 71.9  & 60.2  & 35.7  & 41.9  & 83.6  & 68.4  & 74.1  & 57.4 \\
          & Top20+others & 37.4  & 43.7  & 88.0  & 69.0  & 72.8  & 64.6  & 36.1  & 42.3  & 87.4  & 69.0  & 74.4  & 57.0 \\
          & Top25+others & 37.3  & 43.7  & 88.4  & 71.1  & 72.3  & 60.2  & 36.9  & 43.2  & 86.2  & 70.8  & 75.2  & 62.7 \\
          & Top29+others & 37.9  & 44.2  & 89.9  & 71.5  & 73.0  & 61.4  & 37.1  & 43.5  & 86.8  & 70.4  & 75.8  & 63.2 \\
    \midrule
    \multicolumn{1}{c|}{\multirow{4}[2]{*}{\makecell{MSMARCO (ID)\\$\downarrow$\\Imitate V2}}} & Top15+others & 38.0  & 44.1  & 86.2  & 68.9  & 71.4  & 60.1  & 37.3  & 43.5  & 85.3  & 70.1  & 74.8  & 58.1 \\
          & \textbf{Top20+others} & \textbf{38.4}  & \textbf{44.5}  & \textbf{88.4}  & \textbf{71.9}  & \textbf{77.2}  & \textbf{66.4}  & 36.5  & 42.7  & 87.6  & 70.6  & 76.5  & 63.2 \\
          & Top25+others & 38.3  & 44.5  & 88.4  & 71.1  & 77.0  & 65.3  & 37.3  & 43.6  & 86.8  & 71.6  & 77.2  & 63.4 \\
          & Top29+others & 38.3  & 44.5  & 88.3  & 71.9  & 76.5  & 64.0  & 37.2  & 43.5  & 89.5  & 71.4  & 75.8  & 62.3 \\
    \midrule
    \multicolumn{1}{c|}{\multirow{4}[2]{*}{\makecell{NQ (OOD)\\$\downarrow$\\Imitate V2}}} & Top15+others & 37.4  & 43.6  & 88.0  & 69.5  & 76.0  & 60.4  & 36.0  & 42.1  & 88.3  & 70.5  & 74.4  & 57.5 \\
          & Top20+others & 37.9  & 44.1  & 85.9  & 70.4  & 76.0  & 63.0  & 37.3  & 43.6  & 88.1  & 71.0  & 76.7  & 61.5 \\
          & \textbf{Top25+others} & \textbf{38.3}  & \textbf{44.5}  & \textbf{89.9}  & \textbf{71.3}  & \textbf{77.0 } & \textbf{66.1 } & 36.7  & 42.9  & 88.5  & 69.7  & 74.9  & 60.4 \\
          & Top29+others & 38.0  & 44.2  & 86.6  & 72.3  & 74.0  & 64.0  & 37.0  & 43.3  & 88.8  & 71.7  & 76.5  & 60.7 \\
    \bottomrule
    \end{tabular}%
    }
  \label{tab:sample}%
\end{table*}%

For evaluating the ranking performance on MSMARCO DEV and TREC DL 2019, we adopt \textit{MRR@10} and \textit{NDCG@10}, which are commonly used in information retrieval \citep{nogueira2019passage,hofstatter2020improving,craswell2020overview}.
The Reciprocal Rank (RR) information retrieval measure calculates the reciprocal of the rank at which the first relevant document was retrieved. RR is 1 if a relevant document was retrieved at rank 1. If not, it is 0.5. When averaged across queries, the measure is called the Mean Reciprocal Rank (MRR) \citep{Craswell2009}.
Discounted Cumulated Gain (DCG) assumes that, for a searcher, highly relevant documents are more valuable than marginally relevant documents. It further assumes, that the greater the ranked position of a relevant document is, in terms of any relevance grade, the less valuable it is for the searcher, because the less likely it is that the searcher will ever examine the document and at least has to pay more effort to find it. NDCG is its normalized form that the actual DCG performance for a query is divided by the ideal DCG performance for the same topic, based on the recall base of the topic in a test collection \citep{Craswell2009}.
Following prior works \citep{hofstatter2020improving} about TREC DL Track evaluation, as for MRR, we use a binarization point of 2 for TREC graded relevance labels.
We adopt \textit{AP} and \textit{P@30} following \citep{yang2019simple}, to evaluate the zero-shot ranking performance on TREC MB 2014.

Moreover, we also measure the inter ranking similarity by counting the overlap of the top 10 (\textit{Inter@10}) and Rank Biased Overlap \citep{webber2010similarity} (\textit{RBO@1K}, we set the weighting parameter $p$ to 0.7) between the ranking imitation model and the victim model. \textit{Inter@10} measures the inter ranking similarity by counting the overlap of the top 10 candidates. \textit{RBO@1K} is the Rank Biased Overlap, which calculates the top 1000 weighted ranking overlap between the ranking imitation model and the victim model. The \textit{Inter@10} and \textit{RBO@1K} of \textit{Pairwise BERT (ID)} and all \textit{* Imitate V1} are evaluated with \textit{V1}. The \textit{Inter@10} and \textit{RBO@1K} of all \textit{* Imitate V2} is evaluated with \textit{V2}.



We limit the query-passage pair length to 256 tokens and train our models with a learning rate of $\{1,3,5,7\}\times10^{-6}$ and a batch size of 256.
All the method run on a server configured with 8$\times$32G Tesla V100 and 256G memory.

\subsection{Experimental Results.}

For RQ1, from the results of a comparative study for different sampling strategies in Table \ref{tab:sample}, we can observe that \textit{0.4K} performs relatively better than \textit{1K}. We speculate that \textit{1K} contains more simple triples than \textit{0.4K}, which leads to overfitting of the ranking imitation model and weakens the benefit of hard instances \citep{qu2021rocketqa}. It needs more training efforts and tricks to overcome this problem. Moreover, the adversary prefers \textit{0.4K} because it requires fewer results accesses and is more efficient than \textit{1K}. We leave a thorough investigation to future work. Among sampling strategies in \textit{0.4K}, \textit{top-25 + others} achieves the closest performance to the target model.

For RQ2, the results in Table \ref{tab:sample} also demonstrate the positive effect of the pre-training process of the Pairwise BERT model. The ranking imitation model, i.e., Pairwise BERT, which is firstly pre-trained on MSMARCO (ID) or NQ (OOD) then tuned on $\mathcal{D}$, performs better than the ranking imitation model (Zero Imitate V2). Although \citet{carlini2021extracting} show that it is possible to extract the training data of the language model (GPT-2), extracting the training data of the BERT-based model or even unknown pre-train model is still hard to complete, unless there is a data leak. \textbf{Thus, it is more practical to adopt OOD data (NQ) rather than ID data (a subset of MSMARCO) for the adversary to pre-train the ranking imitation model}. Moreover, \textit{OOD}$\to$\textit{Imitate V2} achieves competitive performance compared to \textit{ID}$\to$\textit{Imitate V2} in terms of the ranking performance and ranking similarity.

As a result, top-25 candidates and randomly selected another 4 candidates (\textit{top-25 + others}) are finally adopted to construct the sampled victims' triples $\mathcal{D}$ (approximately 2.7M). Together with the Pairwise BERT pre-trained on NQ, they are then used to closely imitate the target model.

\begin{table*}[t]
\renewcommand\tabcolsep{2pt}
  \centering
  \caption{Ranking imitation results of performance (\%) comparison with baseline models on MSMARCO DEV, TREC DL 2019, and TREC MB 2014 datasets. \textbf{\textit{Bold}} shows the best ranking imitation performance in terms of \textit{Inter@10} and \textit{RBO@1K}. ``-'' means not applicable.} 
\resizebox{0.8\textwidth}{!}{
  \begin{tabular}{l|cc|cccc|cc}
    \toprule
    \multicolumn{1}{c|}{\multirow{2}[4]{*}{\textbf{Model}}} & \multicolumn{2}{c|}{\textbf{MSMARCO DEV}} & \multicolumn{4}{c|}{\textbf{TREC DL 2019}} & \multicolumn{2}{c}{\textbf{TREC MB 2014}} \\
\cmidrule{2-9}          & \textbf{MRR@10} & \textbf{NDCG@10} & \textbf{MRR@10} & \textbf{NDCG@10} & \textbf{Inter@10} & \textbf{RBO@1K} & \textbf{AP} & \textbf{P@30} \\
    \midrule
    \midrule
    BM25  & 18.7  & 23.4  & 68.5  & 49.7  & -     & -     & 41.4  & 62.6 \\
    TK    & 33.1  & 38.4  & 75.1  & 65.2  & -     & -     & -     & - \\
    BERT-Base & 35.2  & 41.5  & 87.1  & 71.0  & -     & -     & 45.4  & 68.0 \\
    BERT-Large (V1) & 37.1  & 43.3  & 85.5  & 72.6  & -     & -     & 44.9  & 67.4 \\
    {DistilBERT}$_{CAT}$ & 38.2  & 44.2  & 88.9  & 72.8  & -     & -     & 44.6  & 66.5 \\
    MiniLM (V2) & 39.7  & 45.6  & 90.1  & 74.3  & -     & -     & 47.5  & 70.9 \\
    \midrule
    Pairwise BERT (ID) & 34.4  & 40.5  & 88.4  & 71.4  & 75.3  & 64.4  & 41.9  & 65.2 \\
    Imitate ID's Triples & 32.6  & 39.0  & 87.7  & 70.7  & 77.7  & 84.7  & 41.4  & 65.2 \\
    \midrule
    Zero Imitate V1 & 35.7  & 41.7  & 85.7  & 67.0  & 66.1  & 58.6  & 39.5  & 63.1 \\
    ID $\to $ Imitate V1 & 36.2  & 42.4  & 86.8  & 70.2  & \textbf{72.1}  & \textbf{65.2}  & 45.0  & 67.5 \\
    OOD $\to $ Imitate V1 & 36.2  & 42.5  & 84.1  & 69.9  & 71.3  & 63.0  & 42.6  & 65.4 \\
    \midrule
    Zero Imitate V2 & 37.4  & 43.7  & 88.0  & 69.0  & 70.9  & 61.3  & 41.3  & 65.1 \\
    ID $\to $ Imitate V2 & 38.4  & 44.5  & 88.4  & 71.9  & \textbf{77.2}  & \textbf{66.4}  & 45.9  & 68.4 \\
    OOD $\to $ Imitate V2 & 38.3  & 44.5  & 89.9  & 71.3  & 77.0  & 66.1  & 45.2  & 68.2 \\
    \bottomrule
    \end{tabular}%
    }
  \label{tab:extract}%
\end{table*}%

\begin{table}[t]
  \centering
  \caption{Cross ranking similarity results (\%) on OOD dataset for validating the effectiveness of ranking imitation on its target model. \textbf{\textit{Bold}} shows the best ranking imitation performance.}
  \resizebox{0.45\textwidth}{!}{
    \begin{tabular}{l|cc|cc}
    \toprule
    \multicolumn{1}{c|}{\multirow{2}[4]{*}{\textbf{Imitation\textbackslash{}Victim}}} & \multicolumn{2}{c|}{\textbf{V1}} & \multicolumn{2}{c}{\textbf{V2}} \\
\cmidrule{2-5}          & \textbf{Inter@10} & \multicolumn{1}{c}{\textbf{RBO@1K}} & \textbf{Inter@10} & \textbf{RBO@1K} \\
    \midrule
    \midrule
    OOD$\to$Imitate V1 & \textbf{71.3}  & \textbf{63.0}  & 68.4  & 58.2 \\
    \midrule
    OOD$\to$Imitate V2 & 68.6  & 61.3  & \textbf{77.0} & \textbf{66.1} \\
    \bottomrule
    \end{tabular}%
    }
  \label{tab:cross}%
\end{table}%

Then we conduct more comparative imitation ranking experiments on more ranking models and datasets. From Table \ref{tab:extract}, we can observe that:
\begin{itemize}[leftmargin=*]
\item For MSMARCO DEV dataset, ranking imitation models outperform the same architecture (BERT-Base) model, which are trained with data sampled from the ranking list of V1. Also, models imitating V2 achieve approaching or better performance compared with most of baselines. It indicates that, by sampling the ranking results of the victim with better performance, we can train a ranking imitation model with competitive ranking effectiveness results. We suspect it is due to the contribution of knowledge distillation \citep{hofstatter2020improving}.

\item Ranking imitation models based on fine-tuned Pairwise-BERT outperform imitation models only trained on sampled triples, which is also observed in Table \ref{tab:extract}. This indicates that a good start point will benefit the imitation of ranking performance and result similarity, similar to the findings of \citet{krishna2019thieves} on text classification model extraction.

\item As a reference for model similarity, two Pairwise-BERT models trained with same hyperparameters but different random seeds achieve 0.753 Inter@10 and 0.644 RBO@1K. The Pairwise-BERT, trained with same triples labelled by ID (\textit{Imitate ID's Triples}), achieves 0.777 Inter@10 and 0.847 RBO@1K. Thus, we can derive the answer for RQ3 that ranking imitation models with the same architecture, hyperparameters and direct training corpus knowledge (by pseudo triples), especially for the \textit{ID$\to$ Imitate V2}, perform similar with the victim both in ranking effectiveness and agreement. However, from the last six rows of Table \ref{tab:extract} and the cross ranking similarities in Table \ref{tab:cross}, when there is no knowledge of training data, the performance of the victim ranking model has more impact on ranking imitation agreement than the similarity of architecture and hyperparameters. 
\end{itemize}

\section{Ranking Attack Experiments}
In this section, we report the result of ranking attack experiments based on the ranking imitation models and the proposed Pairwise Anchor-based Trigger generation method.

\subsection{Experimental Settings}

\begin{table*}[t]
  \centering
  \caption{White-box attack results for ranking imitation models on TREC DL 2019 dataset. $\boldsymbol{r}$ is the rank of candidates after adding the trigger in front of it.}
    \begin{tabular}{c|l|cccccc}
    \toprule
    \textbf{Target} & \multicolumn{1}{c|}{\textbf{Method}} & \textbf{\% $\boldsymbol{r}\leq20$} & \textbf{\% $\boldsymbol{r}\leq50$} & \textbf{\% $\boldsymbol{r}\leq100$} & \textbf{\% $\boldsymbol{r}\leq500$} & \textbf{ASR} & \textit{avg}. \textit{\textbf{Boost}} \\
    \midrule
    \midrule
    \multirow{4}[2]{*}{\makecell{Pointwise BERT}} & Query+ & 68.4  & 81.8  & 89.8  & 99.0  & 100.0 & 944.7 \\
          & $\mathbf{C}_{aggr}$   & 0.2   & 3.1   & 13.4  & 77.7  & 100.0 & 658.2 \\
          & $\mathbf{C}_{reg}$   & 0.2   & 3.2   & 10.8  & 74.3  & 100.0 & 634.1 \\
          & $\mathbf{C}_{nat}$   & 0.0   & 1.4   & 9.1   & 70.9  & 100.0 & 608.3 \\
    \midrule
    \multirow{7}[1]{*}{\makecell{Imitate V1}} & Query+ & 84.4  & 91.7  & 98.5  & 100.0 & 100.0 & 984.9 \\
          & HotFlip & 0.0   & 0.0   & 0.5   & 12.7  & 68.8  & 173.3 \\
          & PAT   & 6.6   & 14.2  & 22.0  & 48.8  & 96.2  & 471.5 \\
          & \ $\hookrightarrow$greedy & 5.6   & 19.2  & 25.6  & 49.6  & 100.0 & 457.8 \\
          & \quad w/o LM & 9.0   & 18.2  & 26.8  & 56.2  & 98.8  & 530.6 \\
          & \quad w/o NSP & 9.6   & 20.2  & 28.8  & 64.4  & 99.2  & 601.4 \\
          & \quad w/o $Cons.$ & 10.6  & 20.2  & 30.0  & 65.8  & 100.0 & 605.6 \\
    \midrule
    \multirow{7}[1]{*}{\makecell{Imitate V2}} & Query+ & 79.0  & 89.0  & 93.8  & 99.0  & 100.0 & 958.0 \\
          & HotFlip & 0.0   & 0.2   & 0.6   & 13.4  & 69.2  & 150.9 \\
          & PAT   & 9.4   & 15.6  & 25.8  & 51.4  & 96.0  & 485.7 \\
          & \ $\hookrightarrow$greedy & 7.2   & 12.8  & 15.2  & 38.4  & 95.2  & 329.3 \\
          & \quad w/o LM & 13.6  & 23.4  & 34.2  & 64.2  & 98.8  & 603.6 \\
          & \quad w/o NSP & 17.6  & 30.6  & 41.2  & 73.8  & 100.0 & 680.3 \\
          & \quad w/o $Cons.$ & 22.4  & 32.4  & 39.6  & 70.0  & 98.8  & 651.1 \\
    \bottomrule
    \end{tabular}%
  \label{tab:source}%
\end{table*}%

\textbf{Baseline.} 
We compare the following adversarial attack methods, which are originally proposed for text matching or text classification task. Note that in this paper, we mainly investigate those methods based on trigger injection rather than modifying the original passage words or tokens. We modify them for our tasks by keeping the core components:

(1) \textbf{Collisions} \citep{song2020adversarial}. It utilizes gradient-based approaches for generating three types of semantic collisions that are semantically unrelated but judged as relevant by NLP models. As for generating collisions, it does not take the contextual information of the document into consideration. The collision is designed for the pointwise document relevance score model (Birch) \citep{yang2019simple}, we cannot apply the generation method to our ranking imitation model directly. Thus, to compare with this method on the passage ranking task, the same training set from MSMARCO for Pairwise BERT (ID) (2.56M) is adopted to fine-tune a pointwise BERT ranking model, which achieves 87.0 of MRR@10 on TREC DL 2019. Collisions are generated based on this model then transferred to target V1 and V2 model. We adapt the open-sourced code\footnote{https://github.com/csong27/collision-bert} to passage reranking by keeping the core component of collisions unchanged. We abbreviate the aggressive collision baseline as $\mathbf{C}_{aggr}$, regularized aggressive collision baseline as $\mathbf{C}_{reg}$, and natural collision baseline as $\mathbf{C}_{nat}$.

(2) \textbf{HotFlip} \citep{ebrahimi2018hotflip}. It is an universal text attacking method that approximates the effect of replacing a token utilizing the gradient of the target model. Following prior works \citep{wallace2019universal,wallace2020imitation,song2020adversarial} and open-sourced code\footnote{https://github.com/Eric-Wallace/universal-triggers} based on HotFlip attacks, we first initialize the trigger with a sequence of repeating words ``the'', and then iteratively replace all trigger $\boldsymbol{t}$ words. We replace an input token at position $l$ with the token whose embedding minimizes the first-order Taylor approximation of ranking imitation model loss $\mathcal{L}_{R}$:
\begin{align}
    \underset{{\boldsymbol{e}'_l\in \mathcal{V}  }}{\operatorname{argmin}} ({\boldsymbol{e}'_l} - \boldsymbol{e}_l)^\top \nabla_{\boldsymbol{e}_l}\mathcal{L}_{R}(\boldsymbol{q},\boldsymbol{t} \oplus \boldsymbol{p}, \mathcal{A})
    \label{formula:hotflip}
\end{align}
where $\mathcal{V}$ is the model's vocabulary and $\nabla_{\boldsymbol{e}_l}\mathcal{L}_{R}$ is the gradient of $\mathcal{L}_{R}$ with respect to the
input embedding $\boldsymbol{e}$ for the token at position $l$. Considering the imperfection of local first-order approximation, at each iteration, we try all positions $l$ with top 50 tokens \citep{wallace2020imitation} from the formula \ref{formula:hotflip} and choose the replacement with the lowest loss.

Intuitively, inserting the query into the target passage (denotes as \textbf{Query+}) is a straightforward way to boost the rank. Query+ type of attacks usually use the keyword stuffing for spamming. We choose it as one of the baselines and discuss its attack performance and imperceptibility in the following experiments and discussion.

We also test the impact of the components by removing the LM and NSP constraints of PAT (denote as \textbf{w/o LM}, \textbf{w/o NSP}, and \textbf{w/o $\boldsymbol{Cons.}$} respectively) and replacing the beam search with greedy search  (denotes as \textbf{greedy}).

\textbf{Settings.} We choose the ranking imitation models, \textit{ OOD$\to$Imitate V1} and \textit{OOD$\to$Imitate V2}, to generate adversarial triggers using TREC DL 2019 dataset. For simplicity, we denote them as \textit{Imitate V1} and \textit{Imitate V2}, respectively. \textit{BERT-Large (V1)} and \textit{MiniLM-L-12 (V2)} are victims which are adopted to evaluate the effectiveness of these triggers. We attack the target model by adding 6-words triggers into the irrelevant passages, with the hyperparameters of 0.1 and 0.8 for fluency constraint $\lambda_1$ and semantic constraint $\lambda_2$, respectively. Top-200 similar words, selected by the ranking imitation model with the query and vocabulary, are augmented by setting $\epsilon$ to 0.68. The beam search size and temperature are set to 10 and 1.0, respectively. The semantic consistency and fluency described in subsection \ref{subsection:pat} utilize pre-trained BERT NSP model and BERT LM model without fine-tuning on any datasets used in this paper. Top-3 candidates are concatenated as the anchor passage to guide the trigger generation.

For each query, the most irrelevant 5 passages are selected from top-1K that are ranked by the victim models. Then the triggers are inserted at the head of these passages to boost their ranks. We measure the attack results by the rate of successfully boosted candidates (\textbf{ASR}), average boosted ranks (\textit{avg}.\textit{\textbf{Boost}}), and the percentage of the target candidates shifted into top-20 (\textbf{\% $\boldsymbol{r}\leq20$}), top-50  (\textbf{\% $\boldsymbol{r}\leq50$}), top-100 (\textbf{\% $\boldsymbol{r}\leq100$}), and top-500 (\textbf{\% $\boldsymbol{r}\leq500$}). 

\begin{table*}[t]
  \centering
  \caption{Transfer attack results on TREC DL 2019 dataset. \textit{Path} denotes that triggers are generated from the first model and transferred to the second (target) model. $\boldsymbol{r}$ is the rank of candidate passage after adding the trigger in front of it.}
  \renewcommand\tabcolsep{1.pt}
  \resizebox{1\textwidth}{!}{
    \begin{tabular}{l|c|cccccc|c|cccccc}
    \toprule
    \multicolumn{1}{c|}{\textbf{Method}} & \textbf{Path} & \textbf{\% $\boldsymbol{r}\leq20$} & \textbf{\% $\boldsymbol{r}\leq50$} & \textbf{\% $\boldsymbol{r}\leq100$} & \textbf{\% $\boldsymbol{r}\leq500$} & \textbf{ASR} & \textit{avg}. \textit{\textbf{Boost}} & \textbf{Path} & \textbf{\% $\boldsymbol{r}\leq20$} & \textbf{\% $\boldsymbol{r}\leq50$} & \textbf{\% $\boldsymbol{r}\leq100$} & \textbf{\% $\boldsymbol{r}\leq500$} & \textbf{ASR} & \textit{avg}. \textit{\textbf{Boost}} \\
    \midrule
    \midrule
    Query+ & \multicolumn{1}{c|}{\multirow{4}[2]{*}{\makecell{Pointwise\\ BERT\\$\downarrow$\\V1}}} & 74.0  & 83.8  & 91.2  & 98.8  & 100.0 & 949.8 & \multicolumn{1}{c|}{\multirow{4}[2]{*}{\makecell{Pointwise\\ BERT\\$\downarrow$\\V2}}} & 68.4  & 81.8  & 89.8  & 99.0  & 100.0  & 944.7  \\
    $\mathbf{C}_{aggr}$   &       & 0.4   & 1.6   & 1.8   & 11.8  & 73.0  & 116.5  &       & 0.0   & 0.0   & 0.0   & 3.2   & 76.0  & 39.7  \\
    $\mathbf{C}_{reg}$   &       & 0.0   & 0.0   & 0.4   & 5.6   & 65.4  & 58.6  &       & 0.0   & 0.4   & 0.8   & 3.2   & 68.2  & 31.0  \\
    $\mathbf{C}_{nat}$   &       & 3.6   & 6.4   & 8.2   & 30.4  & 94.2  & 298.0  &       & 1.4   & 2.8   & 3.8   & 12.8  & 94.6  & 154.0  \\
    \midrule
    HotFlip & \multicolumn{1}{c|}{\multirow{2}[1]{*}{\makecell{Imitate V2\\$\to$V1}}} & 0.0   & 0.0   & 0.0   & 2.8   & 88.4  & 39.1  & \multicolumn{1}{c|}{\multirow{2}[1]{*}{\makecell{Imitate V1\\$\to$V2}}} & 0.0   & 0.0   & 0.0   & 1.0   & 74.6  & 10.9  \\
    PAT   &       & 7.0   & 13.2  & 19.4  & 36.0  & 47.2  & 316.2  &       & 0.6  & 3.8  & 6.0 & 26.6  & 92.8  & 268.5 \\
    \midrule
    HotFlip & \multicolumn{1}{c|}{\multirow{5}[0]{*}{\makecell{Imitate V1\\$\downarrow$\\V1}}} & 0.0   & 0.0   & 0.2   & 3.8   & 87.4  & 42.6  & \multicolumn{1}{c|}{\multirow{5}[0]{*}{\makecell{Imitate V2\\$\downarrow$\\V2}}} & 0.0   & 0.0   & 0.0   & 1.8   & 75.2  & 12.0  \\
    PAT   &       & 7.2   & 14.4  & 21.6  & 44.6  & 93.8  & 417.3  &       & 4.6   & 10.2  & 15.2  & 39.2  & 92.2  & 373.4  \\
    \ w/o LM &       & 15.6  & 22.2  & 31.2  & 51.6  & 66.2  & 467.0  &       & 4.8   & 11.6  & 20.4  & 50.6  & 97.0  & 471.4  \\
    \ w/o NSP &       & 15.0  & 24.2  & 32.2  & 54.0  & 96.6  & 514.9  &       & 11.0  & 22.0  & 29.4  & 61.6  & 97.8  & 569.1  \\
    \ w/o $Cons.$ &       & 25.2  & 33.6  & 40.0  & 62.6  & 95.6  & 585.4  &       & 17.2  & 27.8  & 33.4  & 59.6  & 97.6  & 557.1  \\
    \bottomrule
    \end{tabular}%
    }
  \label{tab:transfer}%
\end{table*}%

\subsection{Attack Results}
We report the white-box attack results on imitation models in Table \ref{tab:source}. For our imitation models, triggers are able to disorder the candidates and boost their ranks with high success rate and significant boosted rank rate. A small subset of irrelevant candidates can be boosted in the top-20, top-50, and top-100 after inserting collisions. Three types of collisions always beat the HotFlip triggers on different metrics. Aggressive collisions perform better than regularized collisions and natural collisions perform worst among collision baselines.
The trend is consistent with the observation from \citet{song2020adversarial}. 
Adversarial semantic collisions in passage ranking are not as effective as in document ranking. We speculate that the passage ranking model calculates the relevance score by encoding the whole passage into a contextual representation, while the document ranking model \citep{yang2019simple} focuses on key sentences and applies interpolation to calculate the overall relevance, which makes the attack easier.

For our ranking imitation models, triggers are able to disorder the candidates and boost their ranks with high success rate. Triggers generated based on ranking imitation models can boost nearly 20\% to 40\% irrelevant passages into the top-100 ranked by themselves. The \textit{Imitate V1} gets lower attack performance compared with \textit{Imitate V2}.

The PAT without constraints achieves better attack performance for ranking imitation models and victim models. On \textit{Imitate V2}, the full version PAT and the PAT without constraints achieve over 90\% success; triggers generated on PAT without constraints have higher ratio of successfully boosted candidates. The attack success rate of full version PAT is affected by the constraints.
We will analyze the strengths of the constraints in the next subsection.

Moreover, as shown in Table \ref{tab:transfer}, the adversarial ranking attack triggers generated from ranking imitation models with the PAT successfully transfer to victim models. Attacks transfer at a reasonable rate, e.g., the triggers generated on \textit{Imitate V2} with the PAT can transfer to V2 and boost 20.4\% irrelevant passages into top-100.
For the cross transfer attack (\textit{Imitate V1} $\to$ \textit{V2} and \textit{Imitate V2} $\to$ \textit{V1}), the effectiveness of triggers is weakened. It is mainly because their ranking imitation models extract different knowledge that the transferability of triggers across different imitation target models will be affected.

The performance of adversarial semantic collision is significantly weakened when we transfer collisions from \textit{Pointwise BERT} to victim models. It demonstrates that our ranking imitation models are needed for transferable adversarial ranking attacks.

We also utilize the \textit{Pairwise BERT (ID)} to generate triggers, which is only trained on the MSMARCO training set. The triggers generated with PAT against V2 achieve 2.4\% at \textbf{\% $\boldsymbol{r}\leq50$}. It is not as competent as our ranking imitation model, which shows the ranking imitation models considerably enhance the transferability of adversarial triggers.

Although Query+ performs better than PAT in most of cases, it can be detected by anti-SEO solutions \citep{zhou2009osd}. Whereas PAT can find  camouflaged triggers to manipulate embedding-based ranking, by reducing pairwise loss under semantic consistency and fluency constraints. Our evaluations using spamicity detection \citep{zhou2009osd} (refer to the following mitigation analysis by automatic spamicity detection) and human annotation show that the texts generated by Query+ can be easily identified due to the semantic gap between added content and the original passages (also observed by \citet{wu2022prada}), while those produced by PAT are much stealthier. Also, we believe that Query+ will become less effective in ranking manipulation on real-world search engines using DL-based ranking models trained on a large number of query logs.



\subsection{Mitigation Analysis}

\begin{figure}[!t]
\centering
  \includegraphics[width=8.7cm]{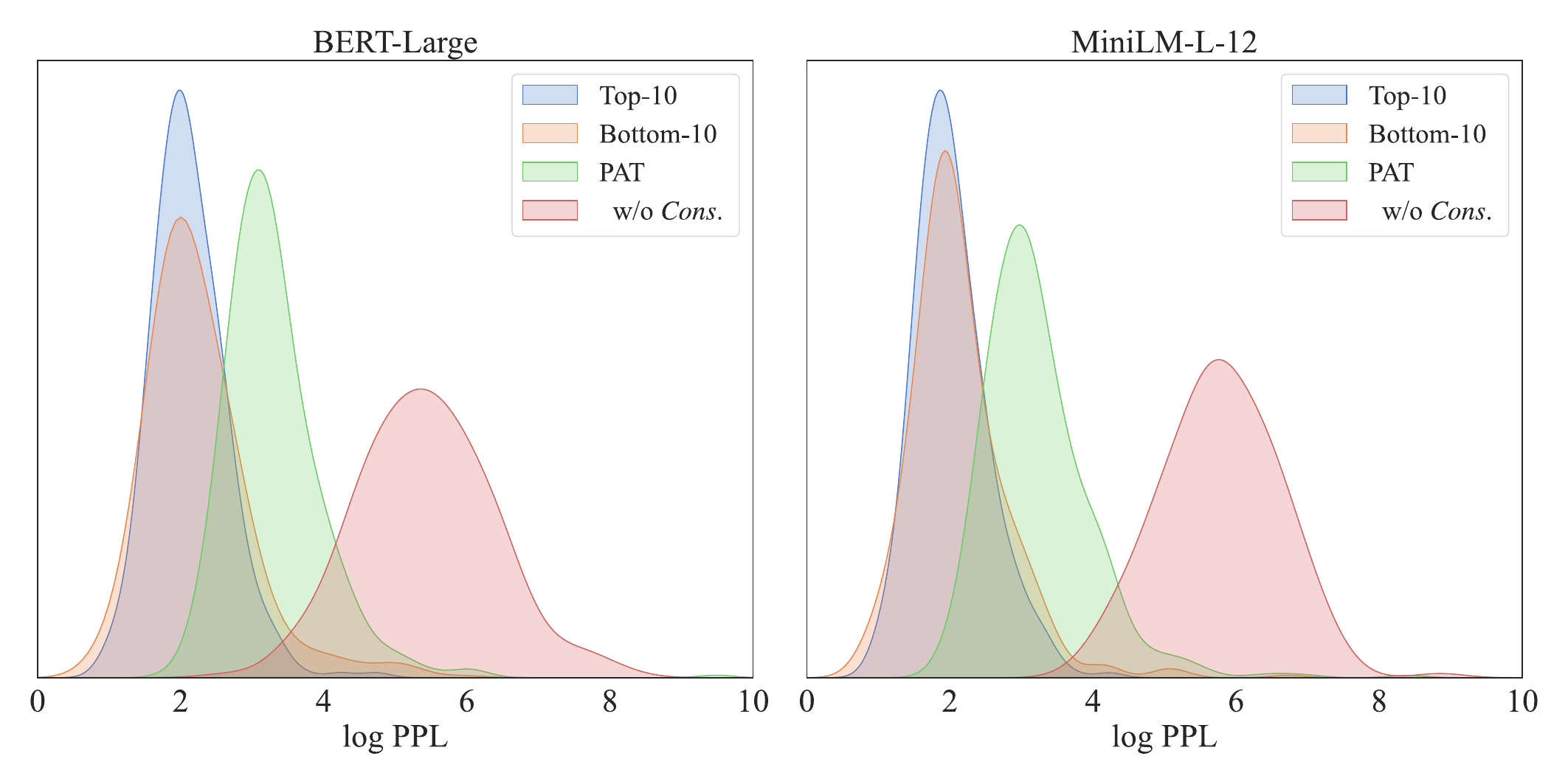}
  \caption{Distributions of log perplexity (PPL) calculated by GPT-2 on TREC DL 2019 passages and triggers.}
  \label{ppl} 
\end{figure}

\textbf{Mitigating by perplexity.} Figure \ref{ppl} shows log perplexity distributions, evaluated by GPT-2 \citep{radford2019language}, on BERT-Large and MiniLM-L-12 for top-10 real passages, bottom-10 real passages, triggers generated by PAT, triggers generated by PAT without LM and NSP constrains (w/o $Cons.$). Note that these triggers are generated based on \textit{Imitate V1} and \textit{Imitate V2} for BERT-Large and MiniLM-L-12, respectively. Since triggers are synthetic, their perplexity is higher than human-generated texts. We can observe a significant gap between the distributions of real passages and triggers generated by PAT w/o $Cons.$. Triggers generated by PAT significantly reduces the perplexity of triggers and have more overlap with the perplexity distribution of real text than triggers generated by PAT w/o $Cons.$. \citet{song2020adversarial} introduced perplexity-based filtering to mitigate the triggers by setting a threshold on language model (LM) perplexity. 
However, as shown in Figure \ref{ppl}, there is a significant distribution overlap between top and bottom passages with PAT triggers. Thresholds that are too strict will result in a higher false positive rate. If the threshold of the filter is set loosely, more attacks will be missed. Thus, coarse-grained perplexity-based filtering cannot effectively mitigate the attack of our proposed PAT triggers.
Moreover, the generated dataset can be used as additional data for increasing the robustness of neural model \citep{li2020bert}. We can further fine-tune the target model using the adversarial triggers to make the model harder to attack. Since adversarial training requires additional optimization processes to make it effective, we will further explore adversarial training-based defense methods in future work.

\begin{table}[t]
  \centering
  \caption{Automatic spamicity detection rates (\%) of Query+ and PAT by tf-idf feature based method.}
    \begin{tabular}{lcccc}
    \toprule
    \textbf{Threshold} & \textbf{0.30} & \textbf{0.20} & \textbf{0.15} & \textbf{0.10} \\
    \midrule
    \midrule
    Query+ & 28    & 80    & 92    & 98 \\
    PAT   & 4     & 14    & 42    & 86 \\
    \bottomrule
    \end{tabular}%
  \label{tab:osd}%
\end{table}%

\textbf{Mitigating by automatic spamicity detection.}
Table \ref{tab:osd} demonstrates the automatic spamicity detection experiment results of Query+ and PAT. The detection method \citep{zhou2009osd} is mainly based on tf-idf feature and has been validated by the Microsoft adCenter. We randomly sample 50 triggers and the corresponding passages. The detection method can computer a spamicity score to the combination of the trigger and the target passage. The table header is the detection threshold. The last two rows are the detection rate (lower is better) under the specific threshold. The detection process can be defined as, given a spamicity threshold and a pair of a passage (may have inserted a trigger) with a query, it determines whether the spamicity score is greater than the threshold. If so, the passage is a suspect of spam content. 
We can observe that PAT performs significantly better than Query+. As the threshold decreases, the detector becomes stricter, and the detection rate of both increases, but this may also lead to more false positives. 

\begin{table}[t]
  \centering
  \caption{Issues counts of triggers generated on ranking imitation model \textit{OOD$\to$Imitate V2} for MiniLM-L-12 by online grammar checkers.}
    \begin{tabular}{l|cc}
    \toprule
    \textbf{Method} & \multicolumn{1}{l}{\textbf{Chegg Writing}} & \multicolumn{1}{l}{\textbf{Grammarly}} \\
    \midrule
    \midrule
    \textbf{Gold}  & 39    & 77  \\
    \midrule
    $\mathbf{C}_{aggr}$   & 68    & 114  \\
    $\mathbf{C}_{reg}$   & 100   & 137  \\
    $\mathbf{C}_{nat}$   & 52    & 88  \\
    Query+ & 71     & 132 \\
    PAT   & 59    & 98  \\
    \quad w/o LM & 67    & 112  \\
    \quad w/o NSP & 57    & 104  \\
    \quad w/o $Cons.$ & 72    & 118  \\
    \bottomrule
    \end{tabular}%
  \label{tab:grammar}%
\end{table}%

\textbf{Mitigating by automatic grammar checker.} \citet{holtzman2019curious} find the degeneration property of the language model, which tends to get low perplexity to repeating tokens. As a result, some disfluent triggers with simply repeating tokens may escape from PPL-based filtering. We employ three online grammar checkers as defense metrics to measure the naturalness of triggers. Specifically, we calculate the average number of errors in the combination of the trigger and the target passage using \textit{Grammarly}\footnote{https://app.grammarly.com/} and \textit{Chegg Writing}\footnote{https://writing.chegg.com/}. For \textit{Grammarly}, we adopt the suggestions numbers about \textit{correctness}, including spelling, grammar, and punctuation. For \textit{Chegg Writing}, we issue numbers about grammar, punctuation, fluency, and typos. 
The issue count results are shown in Table \ref{tab:grammar}. We can observe that, although there is still a gap about the naturalness between the original passage and the trigger, our proposed method and $\mathbf{C}_{nat}$ achieve better naturalness of attacks than other methods. Meanwhile, triggers generated by PAT achieve better attack effectiveness than $\mathbf{C}_{nat}$.

\begin{table}[t]
  \centering
  \caption{Human evaluation results about the trigger generated on ranking imitation model \textit{OOD$\to$Imitate V2} for MiniLM-L-12. }
    \begin{tabular}{l|cc|cc}
    \toprule
    \multicolumn{1}{c|}{\multirow{2}[4]{*}{\textbf{Method}}} & \multicolumn{2}{c|}{\textbf{Imperceptibility}} & \multicolumn{2}{c}{\textbf{Fluency}} \\
\cmidrule{2-5}          & \textbf{avg.} & \textbf{\textit{Kappa}} & \textbf{avg.} & \textbf{\textit{Kendall}} \\
    \midrule
    \midrule
    Query+ & 0.10  & 0.56  & 4.78  & 0.66 \\
    PAT   & 0.70  & 0.62  & 4.35  & 0.76 \\
    \quad w/o LM & 0.14  & 0.67  & 3.18  & 0.90 \\
    \quad w/o NSP & 0.14  & 0.67  & 3.62  & 0.90 \\
    \quad w/o $Cons.$ & 0.12  & 0.63  & 2.68  & 0.85 \\
    \bottomrule
    \end{tabular}%
  \label{tab:human}%
\end{table}%

\textbf{Human evaluation.} To further validate that our attacks are more natural and hard to mitigate than other baselines, we perform a human-subject evaluation about the imperceptibility and the fluency of the trigger. We recruit three annotators to annotate 50 random sampled trigger and the corresponding passage of each ranking attack model. Three annotators are Ph.D. students in information science. Each has over three years of experience in processing/annotating different NLP tasks (e.g., sentiment classification, text summarization, etc.), and received professional training. Note that we keep the random sampling seed consistent with all models. In terms of the \textit{Fluency}, we apply the five-level Likert scale (1-5) to a single trigger. Higher score means the more fluency of the trigger. For \textit{Imperceptibility}, annotators need to assign the 0 (\textit{Perceptible}) or 1 (\textit{Normal}) to the content that a trigger adding in front of the passage. 

A summary of human-subject evaluation statistics, including the annotation consistency test results about \textit{Kappa} value and \textit{Kendall's Tau} coefficient are shown in Table \ref{tab:human}. We can observe that the fluency of the triggers generated by the full version of PAT achieves comparable performance to the query, which demonstrates the effectiveness of the fluency and semantic consistency constraints in PAT. Although \textit{Query+} achieves great fluency and attack performance, it can be easily detected by annotators due to the semantic inconsistency and irrelevance between the query and the passage. Triggers generated by PAT are more imperceptible than other baselines. As a result, human-subject evaluation demonstrates the effectiveness of the semantic consistency and fluency constraints to equip the trigger camouflages.

\begin{table}[t]
  \centering
  \caption{Attack performance of different trigger positions on MiniLM-L-12.}
\renewcommand\tabcolsep{2.pt}
    \begin{tabular}{l|cc|cc|cc}
    \toprule
    \multicolumn{1}{c|}{\multirow{2}[4]{*}{\textbf{Method}}} & \multicolumn{2}{c|}{\textbf{Front}} & \multicolumn{2}{c|}{\textbf{Middle}} & \multicolumn{2}{c}{\textbf{End}} \\
\cmidrule{2-7}          & \textbf{\% $\boldsymbol{r}\leq100$} & \textbf{ASR} & \textbf{\% $\boldsymbol{r}\leq100$} & \textbf{ASR} & \textbf{\% $\boldsymbol{r}\leq100$} & \textbf{ASR} \\
    \midrule
    \midrule
    Query+ & 81.8  & 100.0 & 64.8  & 100.0 & 61.8  & 100.0 \\
    $\mathbf{C}_{nat}$ & 3.8 & 94.6 & 1.8 & 94.0 & 1.8 & 94.2 \\
    PAT   & 15.2  & 92.2  & 9.6   & 93.0  & 17.2  & 99.2 \\
    \quad w/o LM & 20.4  & 97.0  & 10.2  & 96.0  & 18.8  & 99.4 \\
    \quad w/o NSP & 29.4  & 97.8  & 19.8  & 97.6  & 18.4  & 98.4 \\
    \quad w/o $Cons$ & 33.4  & 97.6  & 26.6  & 98.2  & 24.0  & 98.4 \\
    \bottomrule
    \end{tabular}%
  \label{tab:position}%
\end{table}%

\textbf{Position of the trigger injection.} We try to insert triggers at different positions to see how they affect the attack performance. From the experimental results in Table \ref{tab:position}, we can observe that triggers generated from \textit{OOD$\to$Imitate V2} with PAT then transferred to \textit{V2} achieve 15.2\%, 9.6\%, and 17.2\% (\textbf{\% $\boldsymbol{r}\leq50$}) when they are added at the \textit{Front}, \textit{Middle}, and \textit{End} of the passages, respectively. For the other methods, inserting the trigger at the front of the passage usually gets the best performance. However, this may lead to the attack being easily detected. As we have discussed above, the semantic consistency constraint can mitigate this effect.
In this paper, we mainly discuss inserting the trigger at the head of the passage. We will explore more flexible ways to insert triggers and how they will affect the imperceptibility of triggers and the overall semantic in future work.

\begin{figure}[!t]
\centering
  \includegraphics[width=8.5cm]{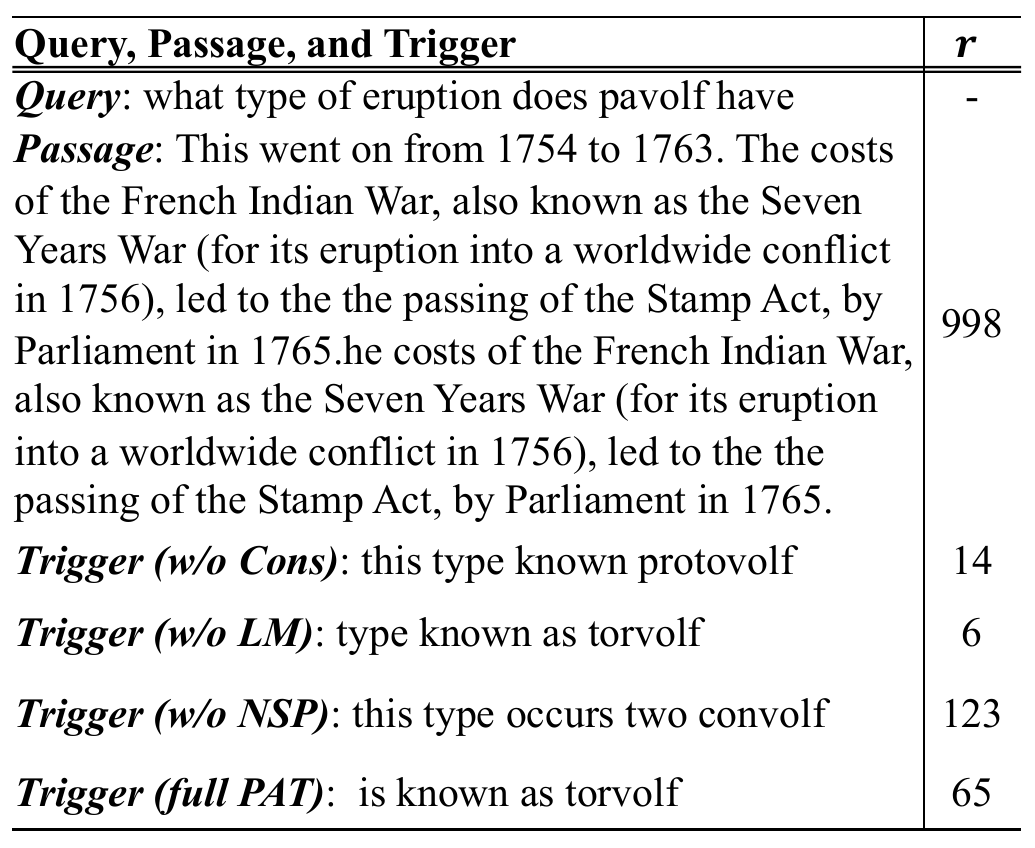}
  \caption{Trigger example generated from \textit{OOD$\to$Imitate V2} for the ranked 998th passage by the  MiniLM-L-12 model with respect to the query. $\boldsymbol{r}$ is the rank of irrelevant passage after adding the trigger in front of it.}
  \label{case} 
\end{figure}

\textbf{Case.} Figure \ref{case} shows example triggers for the query (ID=912070) ``what type of eruption does pavolf have'' from the TREC DL 2019 test set, which are generated from \textit{OOD $\to$ Imitate V2} for the 998th passage ranked by the MiniLM-L-12 model (\textit{V2}). After inserting the trigger, the passage can be ranked as the top candidate. We can also observe that the trigger generated with PAT can be more natural-looking and consistent with the passage content than the trigger generated without the fluency and consistency constraints.



\section{Conclusions}

In this study, we propose an imitation adversarial attack against black-box neural ranking models. The ranking imitation model is able to transparentize the victim ranker via effective pairwise learning. Then, we propose a novel adversarial trigger generator (for each candidate passage) by introducing the pairwise objective function with the anchor content. Based on the competitive ranking effectiveness and similarity between the ranking imitation model and the victim model, adversarial triggers are generated and transferred to the victim ranking model.
The proposed model along with extensive experiment results reveal the vulnerability and risk of black-box text ranking systems.

In the future, we will investigate how to conduct more imperceptible and universal adversarial attacks against text ranking models and detect/defend those attacks effectively.

\section*{Addressing Potential Ethical Concerns}
The goal of our work is to help to make neural ranking models more robust. During performing our research, we used the ACM Ethical Code as a guide to minimize harm. Indeed, the techniques developed in this paper have potential for misuse in terms of attacking existing IR systems with triggers, which has potential negative short-term impacts. However, our intention is not to harm ranking models but instead to publicly release such unintended flaws so that novel defense algorithms can be developed to secure them in the future. This procedure is similar to how white hat hackers expose bugs or vulnerabilities in a software publicly. 
We have demonstrated that adversarial ranking attacks can be accomplished in black-box manner which reveals a greater threat than previous white-box attacks \citep{zhou2020adversarial,song2020adversarial}. This indicates our work \textbf{provides a long-term benefit} to the community and can help to improve IR systems. More importantly, we \textbf{minimized real-world harm} by not exposing any real-world failure or damage to any real users.

\section*{Acknowledgments}
We thank the anonymous reviewers for their valuable comments and suggestions.
This work is supported by the National Natural Science Foundation of China (62106039), the National Key Research and Development Program of China (2019YFB1404702), the Key Research and Development Plan of Zhejiang province (2021C03140), and Alibaba Group through Alibaba Innovative Research Program.

\bibliographystyle{ACM-Reference-Format}
\balance
\bibliography{custom}

\end{document}